\documentclass[twocolumn]{pasj02}
\usepackage[switch, mathlines]{lineno} % add line number to manuscript
\usepackage{natbib}
\usepackage{threeparttable}
\usepackage{xcolor}
\definecolor{xlinkcolor}{rgb}{0, 0, 1}
\usepackage[bookmarks=true, pdfnewwindow=true, colorlinks=true, linkcolor=xlinkcolor, citecolor=xlinkcolor, filecolor=xlinkcolor, urlcolor=xlinkcolor, final=true]{hyperref}
\usepackage{bookmark}

\jyear{2026}
\Received{}%{yyyy/mm/dd}
\Accepted{}%{yyyy/mm/dd}
\graphicspath{{./}} 

\newcommand\chandra{{\it Chandra~}}
\newcommand{\Zsun}{\ensuremath{Z_{\odot}}}

\begin{document} 

%\title{Chemical enrichment in the Ophiuchus cluster core studied by a high-resolution spectroscopy of XRISM}
\title{Metal enrichment in the galaxy group IC 1262}

%%% begin:list of authors
% Do NOT capitalize all letters in "textsc".
\author{
Satish~\textsc{S. Sonkamble},\altaffilmark{1,2}\altemailmark\orcid{0000-0001-8985-8596} \email{satish04apr@gmail.com} 
Dharam~\textsc{V. Lal},\altaffilmark{2}\orcid{0000-0001-5470-305X}
S.~\textsc{Ilani Loubser},\altaffilmark{1,3}\orcid{0000-0002-3937-7126}
and Mahadev~\textsc{B. Pandge}\altaffilmark{4}\orcid{0000-0002-9699-6257}
}
\altaffiltext{1}{Centre for Space Research, North-West University, Potchefstroom 2520, South Africa}
\altaffiltext{2}{National Centre for Radio Astrophysics (NCRA), Tata Institute of Fundamental Research (TIFR), Post Box 3, Ganeshkhind P. O., Pune - 411 007, India}
\altaffiltext{3}{National Institute for Theoretical and Computational Sciences (NITheCS), Potchefstroom 2520, South Africa}
\altaffiltext{4}{Dayanand Science College, Latur - 413 512, Maharashtra, India}
%\footnotetext[$\dag$]{Corresponding author}

%%% end:list of authors

%% !!! Select 3 to 5 words from PASJ's key words !!! 
%% List of Key Words: https://academic.oup.com/pasj/pages/Pasj_Keywords 
%% "\KeyWords{ }" always has to be placed before ``\maketitle'' 

\maketitle

\begin{abstract}
We present a new metal enrichment analysis of a unique galaxy group IC~1262 using archival \chandra and GMRT observations, focusing on metal transport via radio jet, sloshing cold fronts, and shock front. This group shows two sloshing cold fronts along the east and north-west direction which is nearly orthogonal to the north--south orientated radio jet. We report discontinuities in the metallicity at the location of previously detected cold fronts, a more prominent one towards the eastern direction. In addition, the gas inside the cold fronts is 45$\pm$8 per cent more enriched than the gas outside the cold front, suggesting the role of sloshing in transporting metals through the IGrM. We also confirm the presence of a previously reported shock front with higher significance and with greater details. Across this shock, we detect a significant metallicity drop from 0.45$\pm$0.05~\Zsun\ to 0.22$\pm$0.04~\Zsun, located at a projected distance of 78$\pm$2~kpc in the southern direction. The shock could potentially account for the region of gas enrichment seen in the abundance map and profile, which could be the result of a non-Maxwellian electron distribution in its vicinity. This should be considered a contributing factor rather than the sole cause of the observed discontinuity in the abundance. Furthermore, our spectral analysis reveals two temperature X-ray gas preferentially aligned with the radio-jet axis, indicating a possible influence of radio AGN activity on the surrounding gas.
\end{abstract}

%\pagewiselinenumbers 

\section{Introduction \label{sec:intro}}

The intra-cluster medium (ICM) has been extensively studied using the \chandra X-ray observatory, which offers high spatial resolution and sensitivity. One of the first notable findings made possible by the capabilities of \chandra was the detection of distinctive features in the X-ray surface brightness maps of Abell 2142 \citep{2000ApJ...541..542M}, Abell 3667 \citep{2001ApJ...551..160V}, 1E0657-56 \citep{2002ApJ...567L..27M}, and 2A 0335+096 \citep{2003ApJ...596..190M}, characterised by sharp X-ray surface brightness edges. X-ray observations revealed that the gas temperature on the denser side of the edge was colder than that on the less dense side \citep{2004ApJ...616..157F}. Furthermore, electron density profiles across these edges exhibited similar discontinuities, while the pressure profiles remained continuous. Consequently, these observed edges were identified as contact discontinuities between cold, dense gas regions and were named ``cold fronts" \citep{2000ApJ...541..542M}. Subsequently, numerous clusters with cold front features have been observed \citep[e.g.,][]{2001ApJ...555..205M, 2001ApJ...562L.153M, 2007PhR...443....1M, 2017PASJ...69...88I,2019MNRAS.488.2917E, 2019MNRAS.485.1981V, 2019MNRAS.484.4113K}.

A leading interpretation for the origin of cold fronts in dynamically relaxed galaxy clusters or groups is the gas sloshing mechanism. In this scenario, the ICM experiences oscillatory motion within the cluster's gravitational potential well, typically initiated by an off-axis minor merger \citep{2005ApJ...618..227T, 2006ApJ...650..102A}. Numerical simulations support this interpretation, also confirming that such sloshing induced cold fronts can persist for several gigayears \citep{2010ApJ...717..908Z, 2011MNRAS.413.2057R, 2024MNRAS.529..563R}. The detailed studies of both cold fronts and sloshing have mainly focused on dynamics and thermodynamics \citep{2017ApJ...837...34U, 2018NatAs...2..292W, 2019MNRAS.483.1744I, 2020MNRAS.495.2022D}, while less attention has been paid towards the interplay between the sloshing and the chemical properties of groups and clusters (e.g., Abell 496: \citealp{2014A&A...570A.117G}; Centaurus: \citealp{2006MNRAS.371L..65S}; Perseus: \citealp{2011MNRAS.418.2154F}; Abell 2052: \citealp{2010A&A...523A..81D};  Abell 2204: \citealp{2005MNRAS.356.1022S,2009MNRAS.393...71S}; Abell 1201: \citealp{2012ApJ...752..139M}; NGC 5044 \citealp{2014MNRAS.437..730O}; Abell 3581: \citealp{2013MNRAS.435.1108C}; M87: \citealp{2010MNRAS.405...91S}). Abundance studies help us to better understand the cosmic history of metals, their transport and diffusion processes in the ICM \citep{2009PASJ...61S.365S,2010PASJ...62.1445S,2013PASJ...65..111S}. The presence of prominent metallicity peaks in the centres of several cool-core clusters, suggests  that the metals are being produced by the brightest cluster galaxy (BCG) and transported to higher altitudes by radio jets of AGN e.g. the Virgo, Hydra A and M87 clusters. In the study by \citet{2008A&A...482...97S}, it is observed that radio jets are accompanied by metal-enriched X-ray filaments, suggesting that the filaments were carried away by the jets. This phenomenon has been observed in some rich clusters \citep{2005MNRAS.360..133S, 2012ApJ...753...47D} as well as in very few groups, e.g. HCG 62 \citep{2007ApJ...659..275G,2010ApJ...714..758G,2013MNRAS.428...58R}. Since groups are typically 1–2 orders of magnitude less X-ray bright than clusters, they requiring deeper observations and is therefore less studied than clusters.

In a galaxy group, a correlation between the distribution of metals and the position of the cold front \citep[e.g., in NGC 5044:][]{2014MNRAS.437..730O} suggests that the cold front plays an important role in the redistribution of metals in the intragroup medium (IGrM). In addition, observations showed a power-law relationship between radio AGN jet power and the largest projected radius of the metal-enriched gas \citep{2011ApJ...731L..23K,2015MNRAS.452.4361K,2023MNRAS.526..396G}. The cold fronts, gas sloshing, and AGN activity are responsible for metal enrichment in the ICM and IGrM. Other different metal enrichment processes have been suggested in literature like ram pressure stripping \citep{1972ApJ...176....1G}, galactic winds and outflows \citep{2003A&A...408..431M,1980A&A....86...11H,1978ApJ...223...47D}, galaxy galaxy interactions and cluster mergers \citep[e.g.,][]{2022MNRAS.511.3159M}. 

IC~1262 is a rich galaxy group or poor cluster, located at $z = 0.032$ and exhibits complex substructures in its hot gas. These include ripples, a prominent sharp discontinuities extending in both the east and west directions, a large-scale radio jet, recurrent AGN activity, and X-ray cavities filled with radio emission \citep{2019ApJ...870...62P, 2007A&A...463..153T}. In addition, \cite{1999MNRAS.306..857C} detected H$\alpha$+[NII] filaments in the central, brightest galaxy of IC 1262. In this paper, we specifically study the role of metal transport mechanisms via sloshing cold fronts, shock front and radio jets in redistribute metals in the IGrM. To address this, we took advantage of 120~ks \textit{Chandra} observations and low frequency Giant Metrewave Radio Telescope (GMRT) data at 325 MHz of IC~1262.

This paper is structured as follows: We describe the X-ray data analysis, X-ray spectral fitting, and GMRT data analysis in Section~\ref{gmrt-data}. Our results, detection and characterisation of a sloshing cold front and radio jets, including metal abundance and temperature distributions, are presented in Section~\ref{results}. We provide a brief discussion and our findings are summarised in Section~\ref{4}. We adopt H$_0$= 73 km s$^{- 1}$ Mpc$^{- 1}$, $\Omega_M$ = 0.27 and $\Omega_{\Lambda}$ = 0.73 as cosmological parameters. 1 arcsec corresponds to 0.62 kpc at the redshift ($z$ = 0.032) of IC 1262. The error bars quoted in the spectral analysis are at a 90$\%$ confidence limit unless otherwise stated and the metallicities were measured relative to the metallicity table of \cite{1998SSRv...85..161G}.

\section{Observations and data reduction}
\label{gmrt-data}
\subsection{X-ray data}

The field around IC~1262 was observed four times by the \chandra X-ray Observatory, for an effective exposure of 150\,ks (ObsID 6949, 7321, 7322 and 2018) using the ACIS-I and ACIS-S detectors. The observations were reprocessed and Level 2 event files generated of each observation using \textit{{chandra$\textunderscore$repro}} task in CIAO\footnote{\url{http://cxc.harvard.edu/ciao}} 4.9 and employing the latest calibration files (CALDB 4.7.4) provided by the \chandra X-ray center (CXC). We followed the standard \chandra data-reduction threads for the analysis. Periods of high background flares were identified using the \textit{lc$\textunderscore$sigma$\textunderscore$clip} algorithm, with the threshold set at 3$\sigma$ and were removed from further analysis. This yielded a net exposure time of 142 ks. The CIAO \textit{reproject$\textunderscore$obs} script was used to merge data sets by reprojecting event files. X-ray point sources were detected and removed using the \textit{wavdetect} task. The exposure maps in the energy band of 0.5$-$7.0\,keV were generated using the \textit{flux$\textunderscore$obs} script. CIAO scripts \textit{blanksky} and \textit{blanksky$\textunderscore$image} were used to identify appropriate blank sky background files, in proportion to each of the event files. The background events files were reprojected to be consistent with the IC 1262 observations and normalised by the 10-12 keV count rate of the observations. The final image of IC 1262 in the 0.5 - 3.0 keV band was obtained by dividing the merged image by the exposure map to correct for exposure variations and then subtracting the background image of the blank sky.

%%%%%%%%%%%%%%%%%%%%%%%%%%%%%%%%%%%%%%%%%%%%%%%%%%%%%%%%%%%%%%%%%
\begin{figure*}
\center
\includegraphics[scale=0.40]{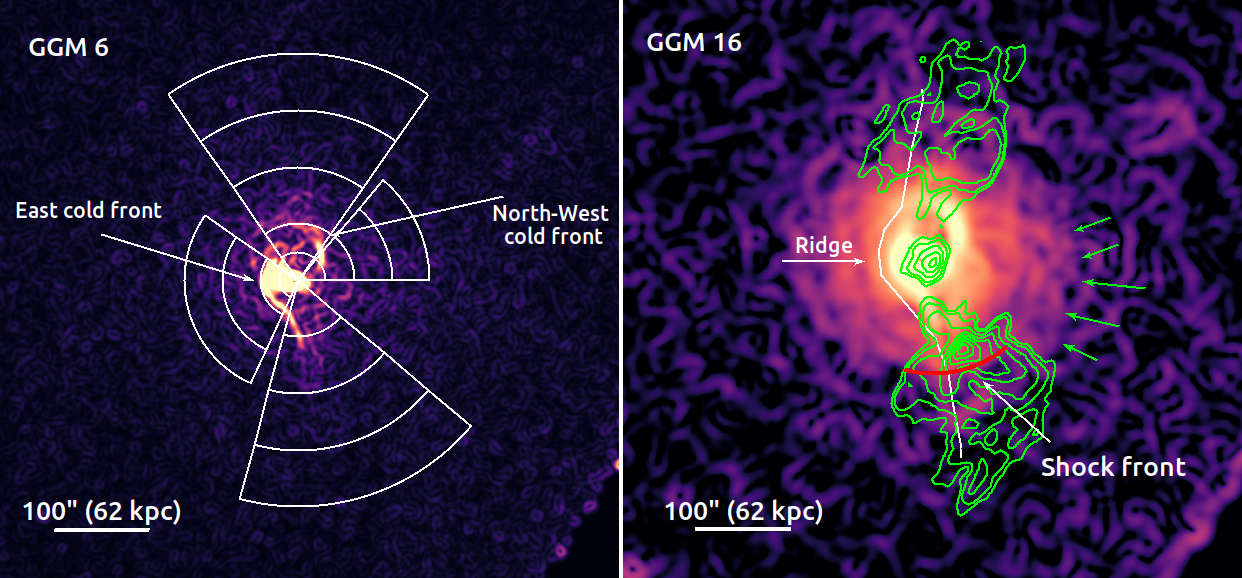}
\caption{{\it left panel:} Background-subtracted and exposure-corrected \chandra\ image in the 0.5--3.0\,keV band, processed with a GGM filter using a kernel of $6\sigma$ pixel (1 pixel = 0.492\arcsec). The eastern and north-western cold fronts are marked, and the regions used to generate radial profile are overlaid. {\it right panel:} GGM-filtered image obtained with a $16\sigma$ kernel. Outer surface brightness edges are indicated by green arrows, while the red arc mark the location of the shock front. The narrow X-ray ridge in the central region identified by \citet{2019ApJ...870...62P}. GMRT 3$\sigma$ radio contours are overlaid in green. North is up and east is to the left. {Alt text: GGM-filtered \chandra\ X-ray images showing cold fronts, shock front, surface brightness edges, central X-ray ridge, and GMRT radio contours at two spatial scales.}}
\label{fig2}
\end{figure*}
%%%%%%%%%%%%%%%%%%%%%%%%%%%%%%%%%%%%%%%%%%%%%%%%%%%%%%%%%%%%%%%%%

The \textit{specextract} task within CIAO was used to generate spectra and the corresponding Redistribution Matrix Files (RMF) and Ancillary Response Files (ARF). The extracted spectra were fitted using the XSPEC package \citep{1996ASPC..101...17A}, version 12.9.1, following the method described in \cite{2015Ap&SS.359...61S}. In the process of fitting the spectra, we incorporate an absorption model (TBABS) to correct for the impact of Galactic absorption.

\subsection{GMRT data}
GMRT archival data (Project code 07MHA01) at 325 MHz with 16 MHz bandwidth was used. Classic AIPS\footnote{\url{https://www.aips.nrao.edu/index.shtml}} was used to analyse the archival GMRT data \citep{1991CuSc...60...95S}. Antennas that did not work and data corrupted due to radio frequency interference were identified and flagged. In order to perform the wide-field imaging particularly important at low GMRT frequencies, we divided the entire field of view of 108$^{\prime}$ into 25 overlapping facets. We further divided the 16 MHz band into 32 sub-bands, each of $\sim$0.5~MHz to correct for the bandwidth smearing. After performing the standard calibration, we imaged the target source. We performed 3--4 rounds of phase-only self-calibrations, and a final amplitude and phase self-calibration were performed in order to correct for antenna dependent phase and amplitude errors. These sets of facets were stitched together for a single image, which was further corrected for the primary beam shape utilising the AIPS task PBCOR in order to achieve rms noise of 0.25 mJy beam$^{-1}$.

\section{Results}
\label{results}

\subsection{X-ray imaging}

To investigate the presence of surface-brightness discontinuities such as cold fronts, shock fronts, or other edge like features associated with the galaxy group IC~1262. We generated Gaussian Gradient Magnitude (GGM; \citealt{2016MNRAS.460.1898S,2022A&A...661A..36S}) maps from the exposure-corrected, background-subtracted \chandra X-ray image in the 0.5–3.0 keV band. Two GGM images were produced using different Gaussian smoothing scales: a 6$\sigma$ kernel to enhance small-scale features in the central regions, and a broader 16$\sigma$ kernel to probe large-scale discontinuities in the outskirts. The resulting GGM maps, shown in Figure~\ref{fig2} (left and right panels), clearly reveal prominent surface-brightness edges extending to the east and north-west of the central dominant (cD) galaxy. The eastern and north-western edges are consistent with cold fronts previously reported in the literature. In addition, we detect faint outer surface-brightness discontinuities toward the east, highlighted by green arrows in Figure~\ref{fig2} (right panel), at a projected distance of $\sim$100 kpc from the group centre. A distinct surface-brightness edge is also identified in the southern direction, which we interpret as a shock front. Furthermore, the GGM image reveals a narrow, elongated X-ray ridge extending over $\sim$200 kpc, roughly aligned with the structure reported by \citet{2007A&A...463..153T} and sketched in Figure~\ref{fig2} (right panel). Motivated by the detection of these surface-brightness discontinuities, in the following sections we investigate the thermodynamical and chemical properties of the IGrM in IC~1262, with particular emphasis on the impact of radio jets, sloshing-induced cold fronts, and the detected shock front on the metal enrichment pattern.

\subsection{Temperature and metallicity maps}
\label{sec3.1}

%%%%%%%%%%%%%%%%%%%%%%%%%%%%%%%%%%%%%%%%%%%%%%%%%%%%%%%%%%%%%%%%%
\begin{figure*}
\center
\includegraphics[scale=0.37]{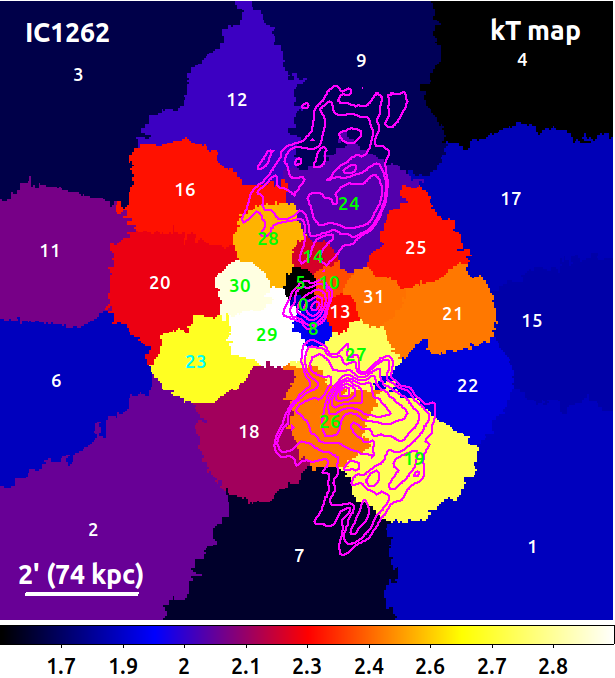}
\hspace{0.2cm}
\includegraphics[scale=0.37]{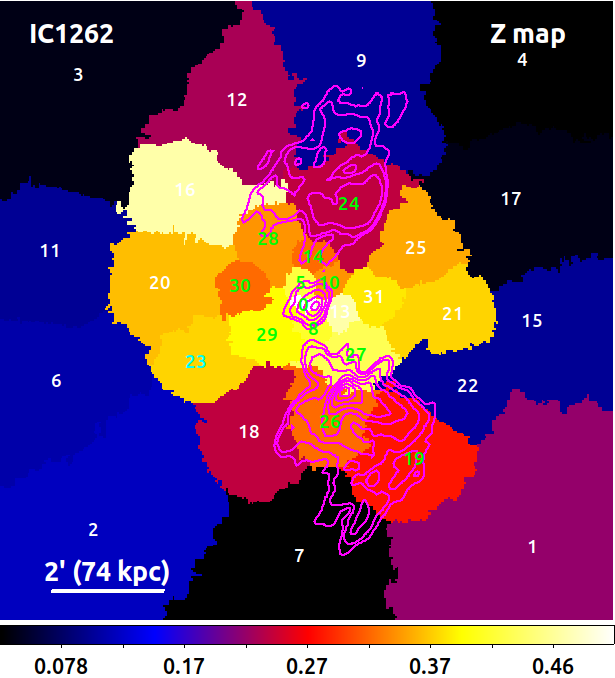}
\caption{{\it left panel:} Projected temperature map in units of keV. {\it right panel:} Projected metallicity map in units of \Zsun\,. The GMRT 325~MHz radio contours at 3$\sigma$ are overlaid in magenta (rms noise $\sim$0.25~mJy~beam$^{-1}$). In both panels, the regions marked with numbers in green are derived from a two-temperature model, while white numbers regions are derived from single temperature model. {Alt text: Temperature and metallicity maps of galaxy group.}}
\label{fig1}
\end{figure*}
%%%%%%%%%%%%%%%%%%%%%%%%%%%%%%%%%%%%%%%%%%%%%%%%%%%%%%%%%%%%%%%%%

%%%%%%%%%%%%%%%%%%%%%%%%%%%%%%%%%%%%%%%%%%%%%%%%%%%%%%%%%%%%%%%%%
\begin{figure*}
\center
\includegraphics[scale=0.53]{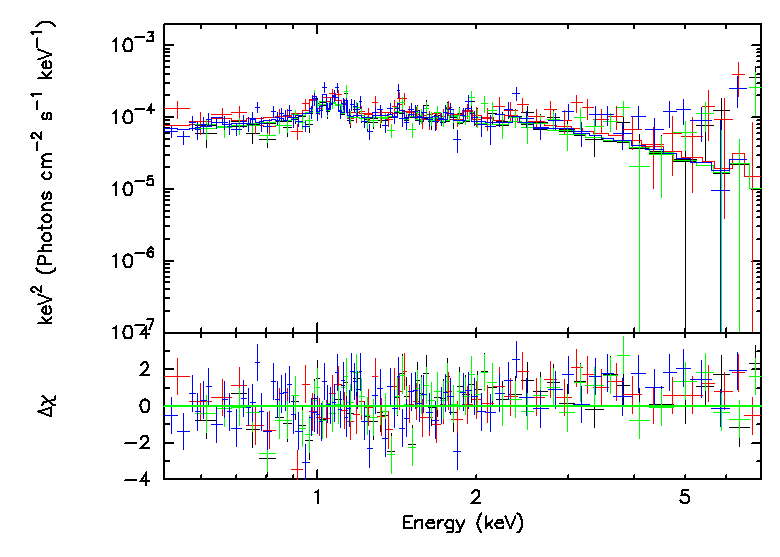}
\includegraphics[scale=0.53]{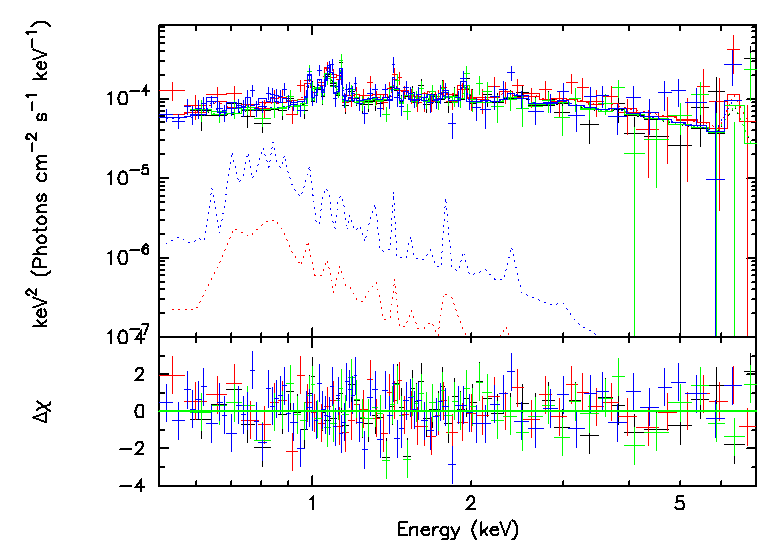}
\vspace{0.2cm}
\caption{{\it top panel:} a typical X-ray spectrum (E$_{2}$ region) obtained from \chandra\,using all the four observations, which shows a poor fit above 4 keV when using a single-temperature model (APEC) model ($\chi^2$/dof = 378.22/271 = 1.39). {\it bottom panel:} the same spectrum is best fitted above 4 keV when a second temperature component (APEC + APEC) is added ($\chi^2$/dof = 300.96/266 = 1.13). {Alt text: X-ray spectrum of galaxy group.}}
\label{spectra}
\end{figure*}
%%%%%%%%%%%%%%%%%%%%%%%%%%%%%%%%%%%%%%%%%%%%%%%%%%%%%%%%%%%%%%%%%

To study the thermal structure and metallicity distribution in the centre of the group, we select a 10\arcmin  $\times$ 10\arcmin\,area around the centre, and generate 2D maps (see Figure~\ref{fig1}) of projected X-ray gas emission using the adaptive binning algorithm, as developed by \citet{2006MNRAS.371..829S}. This algorithm generates spatial bins based on X-ray surface brightness variations in the energy range 0.5--3.0\,keV. We select these regions in such a way that the signal-to-noise (S/N) remains a constant value of 110 (corresponding to $\sim$12100 counts) for each of the bins. We constrain the binning regions by employing the task \textit{constrainfill} and \textit{constrainval} to achieve an approximately spherical shape with \textit{constrainval} = 1. As a result, a total of 32 regions were generated.

We obtain spectra from each region and each ObsID. These spectra are then fitted simultaneously in the energy range of 0.5$-$7\,keV, and the goodness-of-fit described using the $\chi^2$-statistic. We fitted the spectra of all 32 regions with a single temperature model using the TBABS $\times$ APEC. During fitting, Galactic absorption and redshift were fixed at $N_H$ = 2.46$\times10^{20}$ cm$^{-2}$ \citep{1990ARA&A..28..215D}, and 0.0326, respectively. We allowed the temperature and abundance to vary freely and were tied across all ObsIDs. The normalization was permitted to vary independently for each ObsID. We found that for some regions the single temperature (1T) model does not adequately fit the spectrum. Then the fitting was performed with two or multi temperature models as suggested by \cite{2006PASJ...58..719M, 2011PASJ...63S.913K, 2015MNRAS.452.4361K}. This inadequacy of the single temperature model is often seen at the core of groups and clusters, where the temperature structure tends to be more complex. This leads to a poor fit at energies above 4 keV (see Figure~\ref{spectra} top panel) having $\chi^2$/dof = 378.22/271. This issue has been noted in the core of the many objects as well as in some AGN outflow regions which are outside the core \citep{2009MNRAS.396.1449S,2010A&A...523A..81D,2012MNRAS.424.1026T,2002ApJ...580..815M,2007ApJ...658..299O,2018ApJ...853..129K,2003MNRAS.346..525O}. This issue can be solved by adding an extra temperature model (TBABS $\times$ (APEC $+$ APEC)) to the fit and the best fitted spectrum is shown in Figure~\ref{spectra} (bottom panel) with improved $\chi^2$/dof = 300.96/266. It is crucial to note that the single temperature model tends to overestimate the actual emission-weighted temperature in galaxy groups when cool gas (kT $<$ 2–3 keV) is present \citep{1998MNRAS.296..977B}. Analysis of several groups has shown that adding an extra thermal component reduces the calculated kT value by about 10-20\% relative to the single temperature model \citep{2000ApJ...539..172B, 2003MNRAS.340.1375O}. Because, the double temperature model properly accounts for the cool X-ray gas, providing a lower and more physically reasonable value for kT. Metallicity, redshift and Galactic absorption ($N_H$) of the two APEC models are tied to each other and are consistent across all ObsIDs, whereas the temperatures (kT$_{1}$, kT$_{2}$) were permitted to vary and were tied to other ObsIDs. We estimate the best fit-values and their uncertainties at a 68\% confidence level. These values were used as input for the script \textit{paint\_output\_images} \citep{2006MNRAS.371..829S} to produce projected temperature and metallicity maps as shown in Figure~\ref{fig1} along with 325 MHz GMRT 3$\sigma$ radio contours. The uncertainties on each measurement for the temperature map vary from 7\% in the inner regions to 10\% in the outer regions, and for the abundance map they vary from 16\% to 20\%.

We identified 12 regions in the contour bin map (Figure~\ref{fig1}) that are best fitted with a double-temperature (2T) model, shown as green numbers, while the remaining 20 regions are best-fitted with a single-temperature (1T) model shown in white numbers. In particular, 9 of the 12 2T regions are aligned along the radio jet direction, suggesting the presence of multiphase gas associated with radio jet activity. To assess the robustness of these results and to check any model dependent phenomenon, both 1T and 2T models were fitted to all 32 regions independently, and the statistical significance of adding a second temperature component was evaluated using the \textit{simftest} task in XSPEC to compute the $p$-value using 1000 iterations. The 1T model was preferred for $p > 0.01$, while a 2T model was adopted for $p < 0.01$. As summarised in Appendix~\ref{app} (Table~\ref{Tabmap}), out of 9 regions 3 regions 19, 24, and 27 lies along the radio jet have $p$-values close to the significance threshold, likely due to their location in the outer regions of the radio jet where the spectral counts are relatively low. Also, in Region~19, the 1T APEC model yields $\chi^{2}/\mathrm{dof} = 224.83/189 = 1.19 $, while the inclusion of a second thermal component improves the fit to $\chi^{2}/\mathrm{dof} = 98.37/179 = 1.11$. In Region~24, 1T model gives $\chi^{2}/\mathrm{dof} = 305.59/245 = 1.25$, which is reduced to $\chi^{2}/\mathrm{dof} = 283.48/235 = 1.21$ when adopting a 2T model. For Region~27, the 1T fit yields $\chi^{2}/\mathrm{dof} = 237.68/240 = 0.99$, and the fit is marginally improved to $\chi^{2}/\mathrm{dof} = 229.85/234 = 0.98$ with the addition of a second APEC component. The statistical significance of the improvement in these regions were evaluated using the $p$-values of 0.012, 0.021, and 0.025 for Regions~19, 24, and~27, respectively, indicate marginal but consistent improvements in fit quality. These results suggest that a 2T model is statistically favoured in these 3 regions also; however, given the borderline $p$-value significance, the 1T model cannot be conclusively ruled out. Therefore, spatial alignment of the 9 2T regions with the radio jet axis strongly supports a physical association between the thermal structure of the IGrM and the radio AGN activity. This alignment suggests that the observed multi-phase gas is likely produced by jet-ICM interactions. Another Regions 28, 29, 30 lie away from the radio jet but are nevertheless best-fitted with 2T model. This could be due to the eastern cold front and narrow, elongated X-ray ridge reported by \cite{2007A&A...463..153T,2019ApJ...870...62P}. Both cold front and ridge are thought to share a common origin related to gas sloshing and ram pressure striping \citep{2007A&A...463..153T}, processes that can also lead to metal redistribution.

The overall temperature distribution of IC~1262, as evident from the temperature map, is complex and is well correlated with the temperature structure in a typical cool-core cluster, where the temperature in the central region is low and increases gradually as we move outward. The temperature map shows the north radio lobe is cooler than the south lobe. The region along the east cold front is hotter relative to the region along the north-west cold front. The increased metallicity observed along the east and north-west of the central AGN might be attributed to the displacement of metal-rich gas by the traversing cold fronts. However, for the north-western cold front, we cannot confidently confirm that the metallicity enrichment is solely due to the north-western cold front, given its partial overlap with the northern radio lobe, which could possibly uplift metals via entrainment. Overall, the east region appears to be more metal rich than the other parts of the group. As we see the east region ($\sim$150 kpc long) and X-ray ridge ($\sim$180 kpc long) is co-spatial, so we infer that the high metallicity may be associated with X-ray ridge, which is shown by long white arc in Figure~\ref{fig2} (right panel). Similarly in Perseus cluster high abundance X-ray ridge was found by \cite{2005MNRAS.360..133S}. In general, the distribution of metallicity seems to be inhomogeneous and asymmetric.

\subsection{Temperature and metallicity profiles}
\label{shock}

%%%%%%%%%%%%%%%%%%%%%%%%%%%%%%%%%%%%%%%%%%%%%%%%%%%%%%%%%%%%%%%%%%%%%%%%%%%%%%%%%
\begin{figure*}
\center
\includegraphics[scale=0.55]{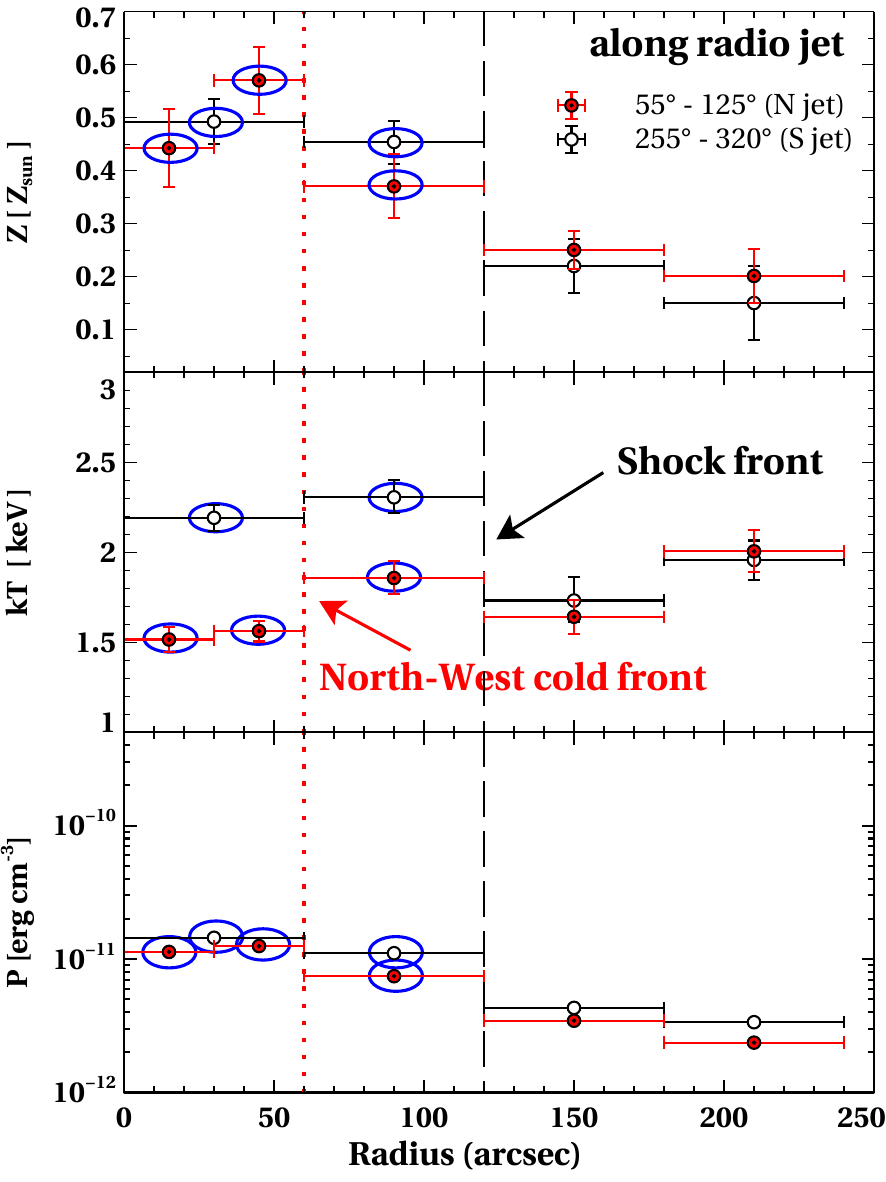}
\includegraphics[scale=0.55]{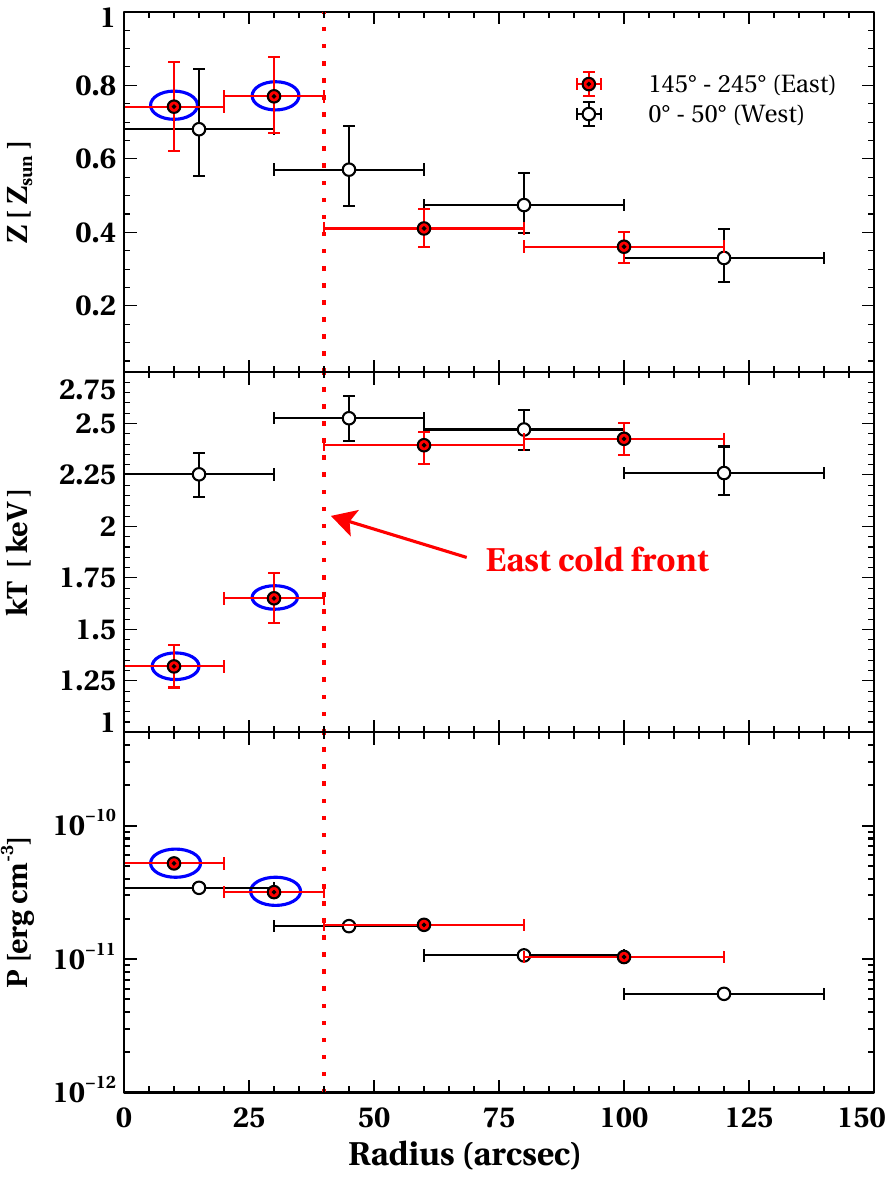}
 \caption{Projected metallicity, temperature and pressure profiles as a function of the radial distance. The blue ellipses overlaid data points are derived from 2T model fitting; all remaining points are from 1T model fitting. The positions of the two cold fronts at 60$^{\prime\prime}$ ($\simeq$ 37 kpc, left panel) and 40$^{\prime\prime}$ ($\simeq$ 25 kpc, right panel) are marked with vertical red dotted lines. {\it left panel:} profiles along the north and south radio jet. The long dashed line represent the location of shock front as discussed in Section~\ref{shock_new}. {\it right panel:} profiles towards the East and West direction. {Alt text: Two radial line plots illustrating the variation of thermodynamical quantities with radius.}}
\label{fig3}
\end{figure*}
%%%%%%%%%%%%%%%%%%%%%%%%%%%%%%%%%%%%%%%%%%%%%%%%%%%%%%%%%%%%%%%%%%%%%%%%%%%%%%%%%

In order to analyse the metal enrichment either due to the cold fronts and/or due to the radio jet, it is necessary to know the radial profiles of temperature, metallicity and electron pressure along the radio jet and along the cold fronts. Therefore, we derived the two projected profiles of IC~1262, one along the radio jet (north-south axis) and another along the off-radio jet, i.e., along the cold fronts (east-west axis). We extracted spectra from X-ray emitting gas along these directions. The widths of the annuli were chosen so that we have at least two annuli before and after the cold front with sufficient S/N ratios, thereby increasing the significance of our results (see Figure~\ref{fig2} left panel). Source spectra, background spectra, photon-weighted response files, and effective area files were generated for each annulus using the CIAO task \textit{specextract}. The extracted spectra in the energy range 0.5-7\,keV were imported to XSPEC and initially modelled with an absorbed single temperature thermal plasma model (TBABS $\times$ APEC). If the fit is not adequate, then we fitted a two-temperature model as described in Section~\ref{sec3.1}. The resulting projected radial profiles of metallicity, temperature, and electron pressure were derived by considering the best fitted values from 1T and 2T model, following the method described in \cite{2024JApA...45...23S} and the resultant plot is shown in Figure~\ref{fig3}. The location of the cold fronts are highlighted by two vertical dotted lines and corresponds to discontinuities in the metallicity, temperature, and pressure profiles. Using the $p$-value criterion, we found that in the south and east directions the inner two annular regions are best described by the 2T model, while in the north direction the inner three annuli prefer the 2T model (see Figure~\ref{fig3}, blue elliptical data points). In contrast, the outer annulus in the south and east, and the outer two annuli in the north, are adequately fitted by the 1T model. For the case of region South 2 (S$_{2}$), the $p$-value is 0.011 it is on the threshold limit, still it remains statistically preferred. In contrast, all four annuli along west are well fitted with the 1T model. All the corresponding values are listed in Table~\ref{Tab3}.

To verify that the temperature discontinuities seen in Figure~\ref{fig3} are not systematic artifacts introduced by the transition between 1T and 2T spectral models, we performed a hard-band ($2$--$7$\,keV) X-ray spectral analysis of regions spanning across the cold front and shock front. This approach is motivated by the study presented in \cite{2000ApJ...539..172B, 2004MNRAS.354...10M}. When dealing with a multiphase plasma, X-ray spectroscopic temperatures are expected to underestimate the gas temperature by $10$--$20$ per cent \citep{1998MNRAS.296..977B, 2000ApJ...539..172B, 2001ApJ...546..100M, 2004MNRAS.354...10M}. For the north-west cold front, no such concern arises: regions N$_{2}$ and N$_{3}$, are both best-fitted by the 2T model (see Table~\ref{Tab3}), so the temperature discontinuity at this location is not affected by any model-transition systematics. However, two locations in Figure~\ref{fig3} coincide with a transition from the 2T model to the 1T model: (i) the east cold front, where region E$_{2}$ (2T) transitions to region E$_{3}$  (1T); and (ii) the shock front, where region S$_{2}$ (2T) transitions to region S$_{3}$ (1T). Since the cooler APEC component in our 2T fits contributes negligibly above $\sim$1.5 keV, as clearly evident from Figure~\ref{spectra} (bottom panel), the cool component (shown by the red dashed curve) falls steeply below this energy. Thus, restricting spectral fits to energies above 2 keV effectively isolates the hot thermal component and renders the recovered temperature largely independent of the choice between the 1T and 2T model \citep{1998MNRAS.296..977B, 2000ApJ...539..172B}. For the east cold front, the hard-band fitting of region E$_{2}$ gives a hot-component temperature fully consistent with the kT$_{1} = 1.65^{+0.12}_{-0.12}$\,keV value in Table~\ref{Tab3} within the 90 per cent confidence interval, while the adjacent outer region E$_{3}$ remains well described by the 1T model at kT = $2.33^{+0.06}_{-0.06}$\,keV. For the shock front, the hard-band fitting of region S$_{2}$ also gets a hot-component temperature consistent with the 2T value of kT$_{1} = 2.30^{+0.15}_{-0.13}$ keV in Table~\ref{Tab3}, while region S$_{3}$ retains its 1T temperature of kT = $1.66^{+0.09}_{-0.08}$\,keV. The persistence of both temperature jumps in the hard-band fits, where the cool component contributes is negligible to the spectral shape, confirms that the discontinuities at the east cold front and the shock front are robust and are not driven by model-dependent systematics arising from the 1T to 2T transition \citep[for more detail see][]{2006PASJ...58..719M, 2015MNRAS.452.4361K}.

\begin{table*}
\caption{Spectral fits for different regions of the radial profile.}
\label{Tab3}
\centering
\footnotesize
\setlength{\tabcolsep}{3pt}
\begin{tabular}{cccccccl}
\hline
Region & Model$^{\#}$ & kT$_{1}$ (keV) & Abundance & kT$_{2}$ (keV)  & Norm $10^{-4} (cm^{-5}$) & $\chi^{2}$/dof & $p$-value \\
\hline
North$_{1}$ & APEC         & $1.49^{+0.07}_{-0.06}$ & $0.44^{+0.07}_{-0.06}$ & \textemdash  & $2.50^{+0.15}_{-0.15}$ & 139.60/146 = 0.96  &   \\
N$_{1}$ & APEC+APEC    & $1.52^{+0.02}_{-0.02}$ & $0.44^{+0.07}_{-0.06}$ & $0.83^{+0.06}_{-0.04}$  & $2.45^{+0.10}_{-0.12}$ & 133.1/126 = 1.05  &  0.000 $^{\dag}$\\
N$_{2}$ & APEC         & $1.59^{+0.02}_{-0.02}$ & $0.31^{+0.02}_{-0.02}$ & \textemdash  & $6.58^{+0.10}_{-0.20}$ & 404.20/374 = 1.08 & \\
N$_{2}$ & APEC+APEC    & $1.56^{+0.05}_{-0.06}$ & $0.57^{+0.06}_{-0.06}$ & $0.99^{+0.05}_{-0.06}$  & $0.42^{+0.10}_{-0.12}$ & 343.36/350 = 0.98 &  0.000$^{\dag}$\\
N$_{3}$ & APEC         & $1.55^{+0.03}_{-0.04}$ & $0.23^{+0.02}_{-0.02}$ & \textemdash  & $5.88^{+0.21}_{-0.20}$ & 341.02/329 = 1.04 &  \\
N$_{3}$ & APEC+APEC    & $1.85^{+0.09}_{-0.08}$ & $0.37^{+0.06}_{-0.05}$ & $1.06^{+0.08}_{-0.07}$  & $0.75^{+0.26}_{-0.24}$ & 313.40/324 = 0.97 &  0.000$^{\dag}$\\
N$_{4}$ & APEC         & $1.64^{+0.09}_{-0.09}$ & $0.25^{+0.03}_{-0.03}$ & \textemdash  & $3.16^{+0.17}_{-0.16}$ & 288.01/255 = 1.13 &  \\
N$_{5}$ & APEC         & $2.00^{+0.11}_{-0.10}$ & $0.20^{+0.05}_{-0.05}$ & \textemdash  & $1.93^{+0.12}_{-0.11}$ & 202.32/212 = 0.95 &  \\
\hline
South$_{1}$ & APEC         & $1.94^{+0.04}_{-0.05}$ & $0.43^{+0.05}_{-0.05}$ & \textemdash  & $4.04^{+0.17}_{-0.16}$ & 382.2/309 = 1.24 &  \\
S$_{1}$ & APEC+APEC    & $2.19^{+0.07}_{-0.07}$ & $0.49^{+0.04}_{-0.05}$ & $1.33^{+0.14}_{-0.12}$  & $3.10^{+0.30}_{-0.50}$ &365.01/304 = 1.20 &  0.007 $^{\dag}$\\
S$_{2}$ & APEC         & $1.91^{+0.06}_{-0.04}$ & $0.39^{+0.05}_{-0.04}$ & \textemdash  & $3.51^{+0.15}_{-0.15}$ & 245.60/282 = 0.87 & \\
S$_{2}$ & APEC+APEC    & $2.30^{+0.15}_{-0.13}$ & $0.45^{+0.04}_{-0.04}$ & $1.44^{+0.25}_{-0.17}$  & $1.10^{+0.08}_{-0.10}$ & 232.05/277 = 0.84 & 0.011 $^{\dag}$\\
S$_{3}$ & APEC         & $1.66^{+0.09}_{-0.08}$ & $0.22^{+0.04}_{-0.03}$ & \textemdash  & $2.82^{+0.15}_{-0.14}$ & 212.93/195 = 1.09 &  \\
S$_{3}$ & APEC+APEC    & $1.61^{+0.11}_{-0.10}$ & $0.22^{+0.05}_{-0.04}$ & $1.10^{+0.17}_{-0.11}$  & $2.18^{+0.20}_{-0.20}$ & 202.87/190 = 1.06 & 0.060 \\
S$_{4}$ & APEC         & $1.95^{+0.10}_{-0.10}$ & $0.15^{+0.07}_{-0.07}$ & \textemdash  & $4.07^{+0.11}_{-0.10}$ & 182.04/178 = 1.02 &  \\
\hline
East$_{1}$ & APEC         & $1.26^{+0.01}_{-0.02}$ & $0.23^{+0.02}_{-0.02}$ & \textemdash  & $2.64^{+0.13}_{-0.13}$ & 242.27/188 = 1.29 &   \\
E$_{1}$ & APEC+APEC    & $1.32^{+0.10}_{-0.10}$ & $0.74^{+0.12}_{-0.13}$ & $1.00^{+0.04}_{-0.03}$  & $0.46^{+0.04}_{-0.03}$ & 179.37/183 = 0.98 &  0.000 $^{\dag}$\\
E$_{2}$ & APEC         & $1.60^{+0.02}_{-0.02}$ & $0.44^{+0.04}_{-0.03}$ & \textemdash  & $3.53^{+0.14}_{-0.14}$ & 378.22/271 = 1.39 &  \\
E$_{2}$ & APEC+APEC    & $1.65^{+0.12}_{-0.12}$ & $0.77^{+0.11}_{-0.10}$ & $1.25^{+0.06}_{-0.05}$  & $0.67^{+0.33}_{-0.21}$ & 300.96/266 = 1.13 &  0.000 $^{\dag}$\\
E$_{3}$ & APEC         & $2.33^{+0.06}_{-0.06}$ & $0.38^{+0.04}_{-0.04}$ & \textemdash  & $5.29^{+0.18}_{-0.18}$ & 286.22/370 = 0.77 &  \\
E$_{3}$ & APEC+APEC    & $2.39^{+0.09}_{-0.06}$ & $0.41^{+0.05}_{-0.06}$ & $0.99^{+0.19}_{-0.20}$  & $0.52^{+0.10}_{-0.10}$ & 281.06/365 = 0.77 &  0.404 \\
E$_{4}$ & APEC         & $2.42^{+0.08}_{-0.07}$ & $0.36^{+0.05}_{-0.04}$ & \textemdash  & $4.20^{+0.15}_{-0.16}$ & 371.18/332 = 1.11 &   \\
\hline
West$_{1}$ & APEC         & $2.25^{+0.12}_{-0.11}$ & $0.68^{+0.12}_{-0.12}$ & \textemdash  & $1.07^{+0.09}_{-0.08}$ & 82.12/101 = 0.81 &   \\
%W$_{1}$ & APEC+APEC    & $1.98^{+0.12}_{-0.13}$ & $0.38^{+0.11}_{-0.09}$ & $0.61^{+0.20}_{-0.24}$  & $2.45^{+0.14}_{-0.13}$ & 113.6/106 = 0.00 &  $^{\dag}$\\
W$_{2}$ & APEC         & $2.52^{+0.11}_{-0.11}$ & $0.57^{+0.11}_{-0.09}$ & \textemdash  & $1.57^{+0.10}_{-0.09}$ & 202.40/149 = 1.36 &  \\
%W$_{2}$ & APEC+APEC    & $2.38^{+0.22}_{-0.21}$ & $0.57^{+0.25}_{-0.16}$ & $0.28^{+0.09}_{-0.08}$  & $2.45^{+0.14}_{-0.13}$ & 123.7/99 = 0.00 &  $^{\dag}$\\
W$_{3}$ & APEC         & $2.47^{+0.10}_{-0.09}$ & $0.47^{+0.08}_{-0.07}$ & \textemdash     & $2.46^{+0.13}_{-0.12}$ & 194.04/189 = 1.02 &   \\
W$_{4}$ & APEC         & $2.25^{+0.11}_{-0.10}$ & $0.33^{+0.08}_{-0.06}$ & \textemdash     & $1.80^{+0.12}_{-0.11}$ & 168.63/150 = 1.12 &  \\
\hline
\multicolumn{6}{l}{$^{\#}$ All models include photoelectric absorption TBABS model. $^{\dag}$ best fitted with 2T model.}
\end{tabular}
\end{table*}

A comparison of the left and right panels of Figure~\ref{fig3} reveals prominent metal enrichment along the direction of the cold fronts. The metallicity profile measured along the northern radio jet and the partial north-west cold front exhibits a steep decline from 0.57$\pm$0.06 \Zsun\,to 0.37$\pm$0.06 \Zsun\, at $\sim$60$^{\prime\prime}$ ($\simeq$ 37 kpc) (see Figure~\ref{fig3} left panel). This metallicity drop is likely a combined effect of the north radio jet and the north-west cold front (see red data points). We also detected a large metallicity drop (from 0.45$\pm$0.05 \Zsun\,to 0.22$\pm$0.04 \Zsun) along the southern radio jet, located at $\sim$120$^{\prime\prime}$ ($\simeq$ 74 kpc). Interestingly, a shock front at the same location was identified by \citet{2009ApJ...693.1142S}, and is highlighted by the vertical long dashed line. More details about the possible origin of the high metallicity associated with the shock front can be found in Section~\ref{4}. Furthermore, the profile of metallicity along the eastern cold front shows a sharp drop in metallicity (from 0.77$\pm$0.10 \Zsun\,to 0.22$\pm$0.05 \Zsun) at $\sim$40$^{\prime\prime}$ ($\simeq$25 kpc) (see Figure~\ref{fig3} right panel). The spatial association between the eastern cold front and location of the metallicity drop, as well as the high metallicity before the cold front and flat profile behind the cold front, suggests that this drop in metallicity may be due to the passage of eastern cold front through the IGrM. This association is also observed in the Perseus cluster \citep{2005MNRAS.360..133S} and M87 cluster \citep{2010MNRAS.405...91S}. To further support our results, we generated a profile from sector $0^\circ$ to $50^\circ$, free from any sloshing or jet-induced asymmetries and discontinuities, for the purpose of comparison with the other profiles. In this context, we observe a steady decrease in metallicity in the outward direction, a typical characteristic associated with a cool core group or cluster. This is opposite to the flat metallicity gradient observed. Thus, the enhancement in metallicity could be attributed to the passage of the eastern cold front, rather than what is typically observed in a cool-core system.

The temperature profile along the southern radio jet is constant until $\sim$120$^{\prime\prime}$ ($\simeq$ 74 kpc), after which it shows a drop from 2.30 $\pm$0.15 keV to 1.61 $\pm$0.10 keV, possibly due to the presence of the shock front reported, which is highlighted by the long dashed black line in the left panel of Figure~\ref{fig3}. We see a hint of a temperature discontinuity from 1.56 $\pm$ 0.05 keV to 1.85 $\pm$0.09 keV near the location of the north-west cold front. This discontinuity is not obvious because the annuli used to extract the profile do not completely cover the north-west cold front. Similarly, for the eastern cold front, at $\sim$40$^{\prime\prime}$ ($\simeq$25 kpc), we see a significant increase in the temperature jump from 1.65 $\pm$0.12 keV to 2.39 $\pm$0.08 keV.

\subsection{Shock influence on metal enrichment}
\label{shock_new}

%%%%%%%%%%%%%%%%%%%%%%%%%%%%%%%%%%%%%%%%%%%%%%%%%%%%%%%%%%%%%%%%%
\begin{figure*}
\center
\includegraphics[scale=0.6]{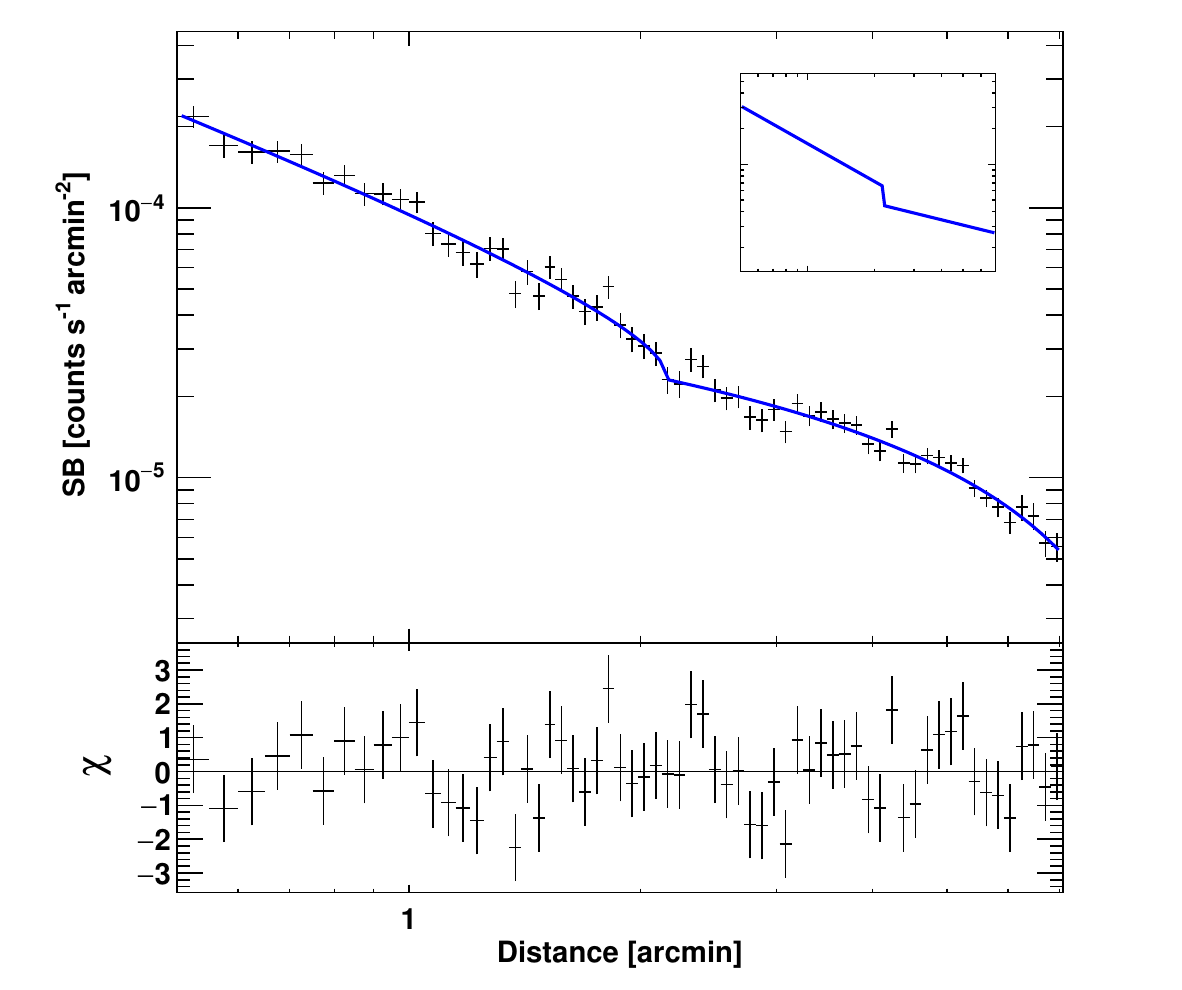}
\caption{X-ray surface brightness profile in the 0.5-3.0 keV energy band extracted along south direction of IC 1262. The power-law density model is shown in inset. {Alt text: Line graph with best-fit model.}}
\label{proffit_shock}
\end{figure*}
%%%%%%%%%%%%%%%%%%%%%%%%%%%%%%%%%%%%%%%%%%%%%%%%%%%%%%%%%%%%%%%%%

Both \cite{2003ApJ...595L...1H} and \cite{2009ApJ...693.1142S} mention the shock front, but did not study it, or the connection with metal redistribution, in detail. From the temperature profile in Figure~\ref{fig3}, we see a clear drop along south direction that could be due to a shock front between 60\arcsec - 120\arcsec. In order to confirm the shock front and its exact location along this direction we, extracted the surface brightness profile up to a radius of 7\arcmin\, from the exposure corrected, background subtracted and vignetting corrected image in the energy range of 0.3-3.0 keV. We then extracted a surface brightness profile and fitted it with a deprojected broken power-law density model using the PROFFIT V 1.5 package
\citep{2011A&A...526A..79E, 2012A&A...541A..57E}. The density model is defined as:

\begin{equation}
n(r) = 
\left\{
\begin{array}{ll}
C n_0 \left( \frac{r}{r_{shock}} \right)^{-\alpha_1},  & if \,  r \leq r_{shock} \\
n_0 \left( \frac{r}{r_{shock}} \right)^{-\alpha_2},    & if \,  r > r_{shock}
\end{array}
\right.
\end{equation}
Here, $n$ is the electron number density, $n_0$ is the density normalization, and $C = n_{e_2}/n_{e_1}$ is the density compression factor of the shock.  $\alpha_1$ and $\alpha_2$ are the power-law indices, $r$ is the radius from the centre of the sector, and $r_{shock}$ is the radius corresponding to the putative shock front. All the parameters of the model were kept free during the fit. The resultant surface brightness profile is shown in Figure~\ref{proffit_shock} and best-fit parameters in Table~\ref{jump}. This profile reveals a significant discontinuity at $\sim$120\arcsec\,(74.4 kpc), accompanied by a corresponding density variation (see the inset image in Figure~\ref{proffit_shock}). Additionally, a pressure discontinuity is observed at this location, indicating the presence of a shock front. From the Rankine-Hugoniot jump conditions, assuming an adiabatic index of $\gamma=5/3$ and the relation given by \cite{2012PASJ...64...49A},

\begin{equation}
\frac{1}{C} = \frac{3}{{4\cal M}^2} + \frac{1}{4},
\end{equation}
the Mach number of the shock is calculated as ${\cal M}$ = 1.38$\pm$0.16.

As the projected temperature profile along the southern direction revealed a decline in temperature after a sharp rise at approximately $120''$, we can estimate the Mach number $\cal M$ of the shock employing the Rankine-Hugoniot condition for the temperature \citep{1959flme.book.....L},

\begin{equation}
\frac{T_2}{T_1} = \frac{5{\cal M}^4+14{\cal M}^2-3}{16{\cal M}^2} \, ,
\label{eq:tjump}
\end{equation}
where $T_1$ and $T_2$ represent the temperature values before and after the shock, respectively. We measure the pre- and post-shock temperature values from the analysis of spectra extracted from wedge-shaped annuli within $60''$--$120''$ and $120''$--$180''$, respectively, along the south direction. This resulted in the temperature ratio:

\begin{equation}
    \frac{T_2}{T_1} = \frac{2.3 \pm 0.20}{1.6 \pm 0.10} = 1.44 \pm 0.15,
\end{equation}
leading to a Mach number of ${\cal M} = 1.45 \pm 0.18$. The Mach number was then used to find the velocity of the shock using the relation ${\cal M} = \frac{v}{c_s}$ \citep{2002ASSL..272....1S}, where $v$ is the velocity of the shock front and $c_s$ is the velocity of sound in the pre-shock gas. The sound speed was calculated using

\begin{equation}
    c_s = \left( \frac{\gamma k_B T}{m_H \mu} \right)^{1/2},
\end{equation}
assuming a monoatomic gas with an adiabatic index $\gamma = 5/3$ and a mean molecular weight $\mu = 0.62$. The pre-shock gas temperature $T_1 = 1.6\pm0.10$ keV gives a sound speed of $c_{s}$ = 641$\pm$20\,km s$^{-1}$, leading to a shock velocity $v_{shock}$ = 930$\pm$119\,km s$^{-1}$. This value is in good agreement with the velocity of 1050 km s$^{-1}$ reported by \cite{2009ApJ...693.1142S}.

Alternatively, the Mach number was also calculated from the observed jump in the pressure values using the relation given by \cite{2010MNRAS.406.1721R}

\begin{equation}
    \frac{P_2}{P_1} = \frac{2\gamma}{\gamma+1} ({\cal M}^2 - 1) + 1.
\end{equation}
The measured projected values of pre- and post-shock pressure are $P_1 = 3.32\pm0.08 \times 10^{-12}$ erg cm$^{-3}$, $P_2 = 7.11\pm0.17 \times 10^{-12}$ erg cm$^{-3}$, respectively, and yielded a Mach number of ${\cal M} = 1.39\pm0.03$. These estimates assume a uniform line-of-sight depth and spherical symmetry for the gas distribution.

Thus, the estimates of the Mach numbers by two independent methods in three different aspects agree within their 1$\sigma$ uncertainties. Using the compression factor, we also calculate the shock velocity $v_{shock} = C \times c_s$ = 988$\pm$140 km s$^{-1}$. The shock velocity estimates obtained using two different methods are in agreement within 1$\sigma$ and are similar to the values reported for other galaxy groups. For example, the shock velocity in NGC 5903 is 710 km s$^{-1}$ \citep{2018MNRAS.473.5248O}, NGC 1132 is 600 km s$^{-1}$ \citep{2018ApJ...853..129K}, while in NGC 2276, it is 865$\pm$120 km s$^{-1}$ \citep{2024A&A...689A..22R}. A comparable velocity of 1100$\pm$66 km s$^{-1}$ has been reported for the NGC 741 group \citep{2024ApJ...976...64R}.

%%%%%%%%%%%%%%%%%%%%%%%%%%%%%%%%%%%%%%%%%%%%%%%%%%%%%%%%%%%%%%%%%%%%%%%%%%%%%
\begin{table*}
\caption{Best-fitting parameters of the broken power-law density model.}
\centering
\small
\begin{tabular}{ccccccccc}\hline
Region &$\alpha_1$ & $\alpha_2$ & $r_{shock}$ & $n_0$ & Compression &$\chi^{2}$/dof & ${\cal M}$   \\
&&&(arcmin)&($10^{-3}$)& (C) &     \\ \hline
South ($255^\circ - 320^\circ$)& $0.95\pm0.13$   &$1.38\pm0.08$  &$2.1\pm0.03$&$5.80\pm0.02$ &$1.54\pm0.22$  &16.85/13&$1.45\pm0.02$  \\  \hline
\end{tabular}
\label{jump}
\end{table*}
%%%%%%%%%%%%%%%%%%%%%%%%%%%%%%%%%%%%%%%%%%%%%%%%%%%%%%%%%%%%%%%%%%%%%%%%%%%%%

\section{Discussion}
\label{4}
The IGrM/ICM is a metal-rich, hot gas found in galaxy groups and clusters, sometimes associated with sloshing cold fronts, radio jets, and shock fronts. This study examines the contribution of these structures to the metal enrichment of the IGrM in IC 1262.

The angular momentum transferred during an off-axis minor merger can trigger gas sloshing, producing spiral- or arc-shaped surface-brightness discontinuities commonly identified as cold fronts \citep{2005ApJ...618..227T,2006ApJ...650..102A,2011MmSAI..82..588J,2011ApJ...743...16Z,2012A&A...544A.103V}. Our analysis reveals a sharp drop in metallicity across the detected discontinuities (cold fronts) in IC~1262, indicating that the sloshing process plays a key role in redistributing metal-enriched gas. Such metallicity contrasts are expected when dense, cooler gas from the core comes into contact with hotter, lower-metallicity gas at larger radii, enhancing discontinuities in density, surface brightness, temperature, and abundance. It is likely that the core of IC~1262 was already metal rich prior to the onset of sloshing, leading to the formation of a pronounced metal discontinuity as the enriched gas was displaced outward. A metallicity drop across a cold front was first reported by \citet{2007ApJ...671..181D}, and similar abundance discontinuities have been observed in several sloshing systems, including Fornax \citep{2017ApJ...851...69S}, Perseus and Abell~1795 \citep{2017MNRAS.468.2506W}, M87 \citep{2010MNRAS.405...91S}, Centaurus \citep{2016MNRAS.457...82S}, Abell~2029 \citep{2013ApJ...773..114P}, and Abell~2204 \citep{2009MNRAS.393...71S}. We find that the metallicity contrast between the gas inside and outside the cold front is large (45$\pm$8 per cent), suggesting the efficiency of sloshing in transporting metals within the IGrM. A comparable behaviour is observed in the abundance profiles of M87 measured with \textit{Suzaku} \citep{2010MNRAS.405...91S}. Numerical simulations and observational studies further support the conclusion that gas sloshing is capable of redistributing significant amounts of metal-enriched material from the central regions to larger radii \citep{2010A&A...523A..81D,2013MNRAS.435.1108C}.

In addition to sloshing, AGN feedback from the jet may contribute to the spatial variations in the metal distribution \citep{2009ApJ...707L..69K, 2011ApJ...731L..23K}.
\cite{2007ApJ...659..275G} investigated a range of possibilities, including entrainment caused by AGN feedback via radio jets or radio lobes. They also considered the effects of expanding X-ray cavities, which can create shock fronts. Additionally, they suggested that ram-pressure stripping from galaxies crossing the group’s centre may contribute. At least in the IC 1262 group, the second generation jet or inner small cavities and their present location and dimensions, as reported by \cite{2019ApJ...870...62P}, are too distant from the shock front for entrainment to take place. However, entrainment due to the first generation jet or large outer cavity south of the centre can not be ruled out. Unfortunately, we do not have sufficient information on the group dynamics, and the system is too complex to disentangle this issue.

Here, we emphasize that in IC 1262 the location of the shock front and the metal-rich gas is co-spatial. \cite{2009AIPC.1201..363S} suggest that the shock may induce turbulence, redistribute metals, and enhance localized abundances within the IGrM/ICM. But other studies by \cite{2001A&A...373.1110P}, \cite{2009A&A...503..373K}, and \cite{2009A&A...508...69P} suggest that in certain scenarios, such as near the boundaries of hot and cold gas and close to shocks, the conventional approximation of Maxwellian electron distributions, which is typically used for computing thermal X-ray spectra, is no longer applicable. In such situations, the existence of non-thermal electrons alters the line emissivities and consequently the X-ray spectrum, thus influencing the metal abundance measurements. This effect in galaxy groups and clusters has been recognized by using data from \textit{Suzaku} X-ray observatory \citep{2010PASJ...62..371H, 2011PASJ...63S1019A}. \cite{2011ApJ...727..126W} demonstrated that non-equilibrium ionization (NEI) effects are largely independent of cluster mass but show a strong dependence on redshift. These effects are particularly pronounced in low-redshift systems like ours, where NEI conditions can enhance emission line ratios, especially near or beneath the shock front by up to an order of magnitude. In addition to NEI, the two-temperature structure of the ICM can further contribute to deviations from ionization equilibrium \citep{2010PASJ...62..335A}. In our analysis, we incorporated a two-temperature model near the detected shock front, located slightly in the outskirts of the system, with a Mach number (${\cal M}$) =  1.45. \cite{2010PASJ...62..335A} showed that in shocks with Mach (${\cal M}$) = 1.5 - 2.0, the electron temperature can be 10–20\% lower than the average ICM temperature for the outskirts of the shock and 30–50\% lower for the shock (${\cal M}$ = 2 - 4) near the core, if the two-temperature structure is taken into account. This reduction in electron temperature can amplify the departure from equilibrium. However, given the location of our shock in the outer region, the influence of this effect is expected to be minimal. Both NEI and two-temperature gas primarily affect line emissivities and ratios (e.g., Fe XXV / Fe XXVI, O VII / O VIII), which can mimic enhanced metallicity when interpreted using equilibrium models. Studies such as \cite{2009A&A...508...69P, 2010A&A...509A..29P} and  \cite{2011ApJ...727..126W} indicate that, in extreme shocks, these effects can lead to apparent abundance overestimates of up to 50–100\%. For mild shocks like the one in our system, the expected bias is more moderate, typically in the range of 10–40\%. Although NEI can alter line ratios and slightly boost the apparent abundance, this study suggests that abundance is a factor of 2 enhanced (from 0.45$\pm$0.05 \Zsun\,to 0.22$\pm$0.04 \Zsun). Therefore, this effect may help explain part of the observed abundance excess but not all of it. The argument of NEI should thus be considered as a contributing factor, not the sole cause of the abundance enhancement. Therefore, further X-ray spectral analysis and comparison with numerical models are necessary to quantify the influence of shock processes on the metal distribution in IGrM/ICM in detail. 

To advance our understanding of metal enrichment processes, future X-ray studies should prioritise high-resolution spatial and spectral analyses of galaxy groups and clusters.

\section{Conclusion}

In summary, this study presents projected temperature profiles that reveal higher temperatures on the X-ray fainter side of each cold front -- contrary to what would be expected in the case of shock fronts. Additionally, we find a sharp drop in metallicity just outside each cold front. This discontinuity in the metal abundance profile is consistent with sloshing activity, which displaces the central, metal-rich gas and brings it into contact with the more metal-poor gas in the outskirts. In this way, sloshing plays a role in transporting metals from the centre outward. The very low metallicity of only $0.20 \pm 0.03$~$Z_{\odot}$ indicates that the gas beyond the cold front represents the untouched outskirts. 

The presence of AGN feedback may also play a role in shaping the metal distribution by inducing spatial variations through the propagation of the radio jet within the IGrM/ICM. We also find that the two temperature gas components are aligned along the direction of the radio jet. This shows close connection between the AGN activity and IGrM. Additionally, we detect a region of enhanced metallicity near the shock front, which may be influenced by non-Maxwellian electron distributions or NEI conditions. While NEI effects can partially account for the observed abundance excess, they should be regarded as a contributing factor rather than the sole explanation for our case. In general, our findings provide valuable insights into the complex physical processes that govern the metal enrichment of galaxy groups and can inform future studies of IGrM/ICM in galaxy groups/clusters.

%%%%%%%%%%%%%%%%%%%%%%%%%%%%%%%%%%%%%%% 

%%%%%%%%%%%%%%%%%%%%%%%%%%%%%%%%%%%%%%%

%\section*{Supplementary data}
%The following supplementary data is available at PASJ online.

\begin{ack}
We thank the referee for his/her comments and suggestions, which have significantly improved the presentation and statistical significance of these results.  SIL is supported in part by the National Research Foundation (NRF) of South Africa (CPRR240414214079). Any opinion, finding, and conclusion or recommendation expressed in this material is that of the author(s), and the NRF does not accept any liability in this regard. MBP gratefully acknowledge the support from the Science and Engineering Research Board (SERB), New Delhi, under the `SERB CRG’ funding with sanction no. CRG/2023/003463. The authors thank the GMRT staff who made the observations possible. The GMRT is operated by the National Centre for Radio Astrophysics of the Tata Institute of Fundamental Research. The data for this work were obtained from the \chandra Data Archive, High Energy Astrophysics Science Archive Research Center (HEASARC). This work has made use of software packages CIAO and Sherpa provided by the \chandra\, X-ray Center. We acknowledge usage of the HyperLeda database, and the NASA/IPAC Extragalactic Database (NED), which is operated by the Jet Propulsion Laboratory, California Institute of Technology, under contract with NASA. 
\end{ack}

% \section*{Funding \label{sec:fund}}
% This work was supported by JSPS KAKENHI grant Nos. 22B00168, 23D06899, and 25C00872 (YF).

% \section*{Data availability \label{sec:data}}
% The raw data underlying the IC 1262 article are publicly available in the \chandra (https://cda.harvard.edu/chaser/) and GMRT archives (https https://naps.ncra.tifr.res.in). Analysed data may be made available at a reasonable request to the corresponding author.

\bibliographystyle{mnras}
\bibliography{mybib_pasj}

\appendix

\section{: Details of the spectral fits for different regions of contbin maps}
\label{app}
Table~\ref{Tabmap} summarises the best fit values and parameters from X-ray spectral fitting using XSPEC.
\begin{table*}
\caption{Spectral fit values for different regions of contbin maps}
\label{Tabmap}
\centering
\scriptsize
\setlength{\tabcolsep}{3pt}
\begin{tabular}{cccccccl}
\hline
Region & Model$^{\#}$ & kT$_{1}$ (keV) & Abundance & kT$_{2}$ (keV)  & Norm $10^{-4} (cm^{-5}$) & $\chi^{2}$/dof & $p$-value \\
\hline
0 & APEC         & $1.73^{+0.03}_{-0.03}$ & $0.51^{+0.03}_{-0.03}$ & \textemdash  & $1.14^{+0.11}_{-0.10}$ & 60.66/70 = 0.87 &  \textemdash \\
  & APEC+APEC    & $1.83^{+0.44}_{-0.40}$ & $0.51^{+0.16}_{-0.11}$ & $1.51^{+0.20}_{-0.16}$  & $0.56^{+0.05}_{-0.06}$ & 60.64/69 = 0.88 & 0.000 $^{\dag}$\\

1 & APEC         & $1.84^{+0.08}_{-0.06}$ & $0.22^{+0.05}_{-0.04}$ & \textemdash  & $5.04^{+0.24}_{-0.26}$ & 299.28/290 = 1.03 &  \textemdash \\
  & APEC+APEC    & $1.88^{+0.24}_{-0.14}$ & $0.24^{+0.05}_{-0.04}$ & $1.87^{+0.20}_{-0.17}$  & $2.45^{+0.14}_{-0.13}$ & 287.79/293 = 1.00 & 0.679\\

2 & APEC         & $2.07^{+0.13}_{-0.12}$ & $0.12^{+0.04}_{-0.03}$ & \textemdash  & $4.42^{+0.22}_{-0.20}$ & 339.55/279 = 1.21 & \textemdash  \\
  & APEC+APEC    & $3.41^{+0.55}_{-0.48}$ & $0.11^{+0.03}_{-0.04}$ & $1.52^{+0.15}_{-0.14}$  & $2.17^{+0.12}_{-0.11}$ & 369.08/302 = 1.22 & 0.033\\

3 & APEC         & $1.68^{+0.13}_{-0.14}$ & $0.04^{+0.03}_{-0.03}$ & \textemdash  & $3.02^{+0.26}_{-0.21}$ & 256.91/218 = 1.18  & \textemdash \\
  & APEC+APEC    & $1.69^{+0.5}_{-0.5}$ & $0.04^{+0.03}_{-0.03}$ & $1.69^{+0.5}_{-0.5}$  &  $1.43^{+0.13}_{-0.11}$ & 308.85/240  = 1.29 & 0.288  \\

4 & APEC         & $1.58^{+0.14}_{-0.13}$ & $0.03^{+0.03}_{-0.02}$ & \textemdash  & $2.61^{+0.18}_{-0.17}$ & 197.81/172 = 1.15 & \textemdash  \\
  & APEC+APEC    & $2.01^{+0.47}_{-0.50}$ & $0.03^{+0.02}_{-0.02}$ & $1.29^{+0.10}_{-0.20}$  & $1.19^{+0.09}_{-0.10}$ & 178.04/169 = 1.05 & 0.057 \\

5 & APEC         & $1.58^{+0.03}_{-0.03}$ & $0.42^{+0.06}_{-0.05}$ & \textemdash  & $2.48^{+0.14}_{-0.14}$ & 136.09/152 = 0.89 & \textemdash \\
  & APEC+APEC    & $2.23^{+0.21}_{-0.20}$ & $0.52^{+0.08}_{-0.07}$ & $1.37^{+0.04}_{-0.03}$  & $1.11^{+0.08}_{-0.07}$ & 126.37/150 = 0.84 & 0.001 $^{\dag}$\\

6 & APEC         & $1.84^{+0.13}_{-0.11}$ & $0.11^{+0.05}_{-0.04}$ & \textemdash  & $2.63^{+0.18}_{-0.19}$  & 182.11/160 = 1.14 & \textemdash \\
  & APEC+APEC    & $1.84^{+0.29}_{-0.22}$ & $0.11^{+0.05}_{-0.03}$ & $1.83^{+0.29}_{-0.22}$  & $1.32^{+0.08}_{-0.09}$ & 182.11/159 = 1.14 & 0.872 \\

7 & APEC         & $1.64^{+0.09}_{-0.08}$ & $0.03^{+0.02}_{-0.01}$ & \textemdash  & $4.94^{+0.27}_{-0.26}$ & 267.47/233 = 1.14 & \textemdash \\
  & APEC+APEC    & $1.62^{+0.31}_{-0.23}$ & $0.03^{+0.02}_{-0.01}$ & $1.63^{+0.30}_{-0.25}$  & $2.48^{+0.14}_{-0.13}$ & 246.29/217 = 1.13 & 0.678\\

8 & APEC         & $1.91^{+0.08}_{-0.05}$ & $0.41^{+0.08}_{-0.06}$ & \textemdash  & $4.37^{+0.22}_{-0.26}$ & 199.97/184 = 1.09 & \textemdash \\
  & APEC+APEC    & $2.68^{+0.26}_{-0.23}$ & $0.37^{+0.06}_{-0.05}$ & $1.50^{+0.06}_{-0.05}$  & $2.16^{+0.01}_{-0.01}$ & 173.65/179 = 0.97 & 0.001 $^{\dag}$ \\

9 & APEC         & $1.70^{+0.09}_{-0.08}$ & $0.10^{+0.03}_{-0.02}$ & \textemdash  &  $2.92^{+0.19}_{-0.18}$ & 238.51/219 = 1.08 & \textemdash \\
  & APEC+APEC    & $1.69^{+0.34}_{-0.25}$ & $0.10^{+0.03}_{-0.02}$ & $1.66^{+0.36}_{-0.22}$  & $1.48^{+0.09}_{-0.08}$ & 207.61/201 = 1.03 & 0.488\\

10 & APEC         & $1.67^{+0.05}_{-0.05}$ & $0.29^{+0.05}_{-0.04}$ & \textemdash  &  $1.38^{+0.09}_{-0.08}$ & 115.10/125 = 0.92 & \textemdash \\
   & APEC+APEC    & $2.38^{+0.24}_{-0.22}$ & $0.34^{+0.06}_{-0.05}$ & $1.39^{+0.05}_{-0.05}$  & $0.64^{+0.05}_{-0.04}$ & 106.92/122 = 0.88 & 0.001 $^{\dag}$\\

11 & APEC         & $2.09^{+0.18}_{-0.19}$ & $0.10^{+0.06}_{-0.05}$ & \textemdash  &  $2.43^{+0.21}_{-0.18}$ & 180.60/154 = 1.17 & \textemdash \\
   & APEC+APEC    & $2.28^{+0.58}_{-0.79}$ & $0.08^{+0.05}_{-0.04}$ & $1.79^{+1.04}_{-0.32}$  & $1.25^{+0.09}_{-0.08}$ & 167.96/144 = 1.16 & 0.197 \\

12 & APEC         & $2.01^{+0.19}_{-0.13}$ & $0.23^{+0.09}_{-0.06}$ & \textemdash  &  $1.95^{+0.15}_{-0.16}$ & 163.22/157 = 1.04 & \textemdash \\
   & APEC+APEC    & $3.12^{+0.62}_{-0.54}$ & $0.20^{+0.07}_{-0.06}$ & $1.49^{+0.14}_{-0.11}$  & $0.97^{+0.08}_{-0.07}$ & 142.42/146 = 0.97 & 0.050 \\

13 & APEC         & $2.29^{+0.10}_{-0.09}$ & $0.47^{+0.10}_{-0.08}$ & \textemdash  &  $2.18^{+0.14}_{-0.13}$ & 116.02/138 = 0.84 & \textemdash \\
   & APEC+APEC    & $2.69^{+0.47}_{-0.75}$ & $0.47^{+0.10}_{-0.09}$ & $2.01^{+1.00}_{-0.21}$  & $1.56^{+0.07}_{-0.06}$ & 115.63/136 = 0.85 & 0.379 \\

14 & APEC         & $1.56^{+0.04}_{-0.05}$ & $0.26^{+0.04}_{-0.04}$ & \textemdash  & $1.90^{+0.12}_{-0.11}$ & 176.16/127 = 1.38 & \textemdash \\
   & APEC+APEC    & $2.22^{+0.25}_{-0.21}$ & $0.32^{+0.06}_{-0.05}$ & $1.32^{+0.04}_{-0.04}$  & $0.86^{+0.06}_{-0.07}$ & 167.85/126 = 1.33 & 0.004 $^{\dag}$\\

15 & APEC         & $1.81^{+0.12}_{-0.10}$ & $0.10^{+0.04}_{-0.03}$ & \textemdash  &  $3.41^{+0.25}_{-0.24}$ & 196.79/177 = 1.11 & \textemdash \\
   & APEC+APEC    & $1.66^{+0.72}_{-0.22}$ & $0.08^{+0.04}_{-0.03}$ & $1.87^{+0.39}_{-0.55}$  & $1.74^{+0.13}_{-0.12}$ & 166.11/164 = 1.01 & 0.284  \\

16 & APEC         & $2.30^{+0.13}_{-0.11}$ & $0.47^{+0.12}_{-0.09}$ & \textemdash  & $2.55^{+0.17}_{-0.17}$  & 192.72/162 = 1.18 & \textemdash \\
   & APEC+APEC    & $2.20^{+0.27}_{-0.17}$ & $0.42^{+0.11}_{-0.09}$ & $2.27^{+0.22}_{-0.22}$  & $1.31^{+0.09}_{-0.09}$ & 175.97/152 = 1.16 & 0.679  \\

17 & APEC         & $1.83^{+0.09}_{-0.08}$ & $0.04^{+0.02}_{-0.02}$ & \textemdash  &  $4.56^{+0.24}_{-0.23}$ & 259.05/263 = 0.98 & \textemdash \\
   & APEC+APEC    & $2.36^{+0.32}_{-0.33}$ & $0.03^{+0.02}_{-0.01}$ & $1.36^{+0.25}_{-0.18}$  & $2.33^{+0.14}_{-0.12}$ & 234.69/245 = 0.96 & 0.043 \\

18 & APEC         & $2.15^{+0.09}_{-0.08}$ & $0.24^{+0.06}_{-0.05}$ & \textemdash  & $4.24^{+0.22}_{-0.21}$ & 219.87/205 = 1.07 & \textemdash \\
   & APEC+APEC    & $2.15^{+0.30}_{-0.24}$ & $0.24^{+0.06}_{-0.05}$ & $2.14^{+0.29}_{-0.26}$  & $2.12^{+0.11}_{-0.11}$ & 210.83/195 = 1.08 & 0.922 \\

19 & APEC         & $1.95^{+0.08}_{-0.09}$ & $0.30^{+0.07}_{-0.06}$ & \textemdash  & $3.12^{+0.20}_{-0.18}$ & 224.83/189 = 1.19  & \textemdash \\
   & APEC+APEC    & $2.74^{+0.42}_{-0.39}$ & $0.28^{+0.07}_{-0.05}$ & $1.53^{+0.11}_{-0.08}$  & $1.54^{+0.10}_{-0.09}$ & 198.37/179 = 1.11 & 0.012  $^{\dag}$\\

20 & APEC         & $2.25^{+0.10}_{-0.09}$ & $0.36^{+0.08}_{-0.07}$ & \textemdash  & $3.44^{+0.19}_{-0.18}$ & 212.93/195 = 1.09 & \textemdash\\
   & APEC+APEC    & $2.26^{+0.39}_{-0.32}$ & $0.36^{+0.07}_{-0.06}$ & $2.23^{+0.30}_{-0.41}$  & $1.72^{+0.09}_{-0.09}$ & 212.93/194 = 1.09 & 0.764 \\

21 & APEC         & $2.44^{+0.12}_{-0.10}$ & $0.37^{+0.09}_{-0.08}$ & \textemdash  & $2.66^{+0.16}_{-0.15}$  & 194.73/172 = 1.13 & \textemdash \\
   & APEC+APEC    & $2.45^{+0.30}_{-0.26}$ & $0.37^{+0.09}_{-0.07}$ & $2.40^{+0.33}_{-0.23}$  & $1.37^{+0.09}_{-0.08}$ & 177.84/163 = 1.09 & 0.708 \\

22 & APEC         & $1.88^{+0.09}_{-0.08}$ & $0.10^{+0.04}_{-0.03}$ & \textemdash  &  $3.73^{+0.22}_{-0.21}$ & 196.42/171 = 1.14 & \textemdash \\
   & APEC+APEC    & $1.88^{+0.38}_{-0.29}$ & $0.10^{+0.04}_{-0.03}$ & $1.84^{+0.40}_{-0.23}$  & $1.88^{+0.11}_{-0.10}$ & 189.85/164 = 1.16 & 0.641  \\

23 & APEC         & $2.67^{+0.15}_{-0.14}$ & $0.37^{+0.10}_{-0.08}$ & \textemdash  & $3.02^{+0.18}_{-0.16}$ & 122.03/157 = 0.77 & \textemdash \\
   & APEC+APEC    & $2.64^{+0.99}_{-0.64}$ & $0.35^{+0.10}_{-0.08}$ & $2.60^{+1.02}_{-0.61}$  & $1.54^{+0.09}_{-0.08}$ & 117.29/151 = 0.77 & 0.726 \\

24 & APEC         & $1.66^{+0.04}_{-0.05}$ & $0.24^{+0.04}_{-0.03}$ & \textemdash  &  $3.75^{+0.19}_{-0.18}$ & 305.59/245 = 1.25 & \textemdash \\
   & APEC+APEC    & $2.04^{+0.18}_{-0.19}$ & $0.24^{+0.04}_{-0.03}$ & $1.42^{+0.08}_{-0.05}$  & $1.84^{+0.10}_{-0.09}$ & 283.48/235 = 1.21 & 0.021  $^{\dag}$\\

25 & APEC         & $2.30^{+0.11}_{-0.10}$ & $0.35^{+0.09}_{-0.07}$ & \textemdash  &  $2.08^{+0.14}_{-0.13}$ & 99.38/145 = 0.68 & \textemdash \\
   & APEC+APEC    & $2.32^{+0.51}_{-0.39}$ & $0.35^{+0.10}_{-0.07}$ & $2.28^{+0.53}_{-0.36}$  & $1.04^{+0.06}_{-0.07}$ & 97.72/141 = 0.69 & 0.814 \\

26 & APEC         & $1.61^{+0.06}_{-0.04}$ & $0.26^{+0.05}_{-0.04}$ & \textemdash  &  $3.83^{+0.20}_{-0.22}$ & 211.94/170 = 1.25 & \textemdash \\
   & APEC+APEC    & $2.44^{+0.29}_{-0.26}$ & $0.32^{+0.05}_{-0.04}$ & $1.36^{+0.04}_{-0.03}$  & $1.73^{+0.10}_{-0.10}$ & 193.06/163 = 1.18 & 0.000  $^{\dag}$\\

27 & APEC         & $2.16^{+0.07}_{-0.06}$ & $0.47^{+0.07}_{-0.06}$ & \textemdash  &  $4.76^{+0.23}_{-0.22}$ & 237.68/240 = 0.99 & \textemdash \\
   & APEC+APEC    & $2.75^{+0.30}_{-0.29}$ & $0.43^{+0.06}_{-0.05}$ & $1.75^{+0.13}_{-0.10}$  & $2.39^{+0.10}_{-0.10}$ & 229.85/234 = 0.98 & 0.025 $^{\dag}$\\

28 & APEC         & $1.75^{+0.04}_{-0.05}$ & $0.33^{+0.04}_{-0.04}$ & \textemdash  &  $4.35^{+0.20}_{-0.20}$ & 257.82/214 = 1.20 & \textemdash \\
   & APEC+APEC    & $2.52^{+0.22}_{-0.20}$ & $0.34^{+0.05}_{-0.04}$ & $1.42^{+0.04}_{-0.04}$  & $2.07^{+0.11}_{-0.10}$ & 237.74/209 = 1.14 & 0.000  $^{\dag}$\\

29 & APEC         & $1.86^{+0.03}_{-0.03}$ & $0.38^{+0.03}_{-0.03}$ & \textemdash  &  $11.45^{+0.33}_{-0.33}$ & 501.38/327 = 1.53 & \textemdash \\
   & APEC+APEC    & $2.97^{+0.17}_{-0.16}$ & $0.39^{+0.04}_{-0.03}$ & $1.44^{+0.03}_{-0.02}$  & $5.41^{+0.17}_{-0.17}$ & 439.62/320 = 1.37 & 0.000 $^{\dag}$\\

30 & APEC         & $2.12^{+0.08}_{-0.07}$ & $0.34^{+0.06}_{-0.05}$ & \textemdash  &  $3.53^{+0.18}_{-0.17}$ & 139.60/165 = 0.85 & \textemdash \\
   & APEC+APEC    & $2.93^{+0.37}_{-0.33}$ & $0.32^{+0.06}_{-0.05}$ & $1.64^{+0.13}_{-0.10}$  & $1.75^{+0.10}_{-0.10}$ & 133.71/161 = 0.83 & 0.011 $^{\dag}$\\

31 & APEC         & $2.43^{+0.10}_{-0.09}$ & $0.38^{+0.08}_{-0.07}$ & \textemdash  &  $2.47^{+0.14}_{-0.13}$ & 163.89/165 = 0.99 & \textemdash \\
   & APEC+APEC    & $2.40^{+0.43}_{-0.32}$ & $0.35^{+0.08}_{-0.07}$ & $2.38^{+0.41}_{-0.33}$  & $1.25^{+0.07}_{-0.07}$ & 161.07/161 = 1.00  & 0.772 \\

\hline
\multicolumn{6}{l}{$^{\#}$ All models include photoelectric absorption TBABS model. $^{\dag}$ best fitted with 2T model.}
\end{tabular}
\end{table*}

\end{document}